\documentclass[apjl]{emulateapj}

\usepackage{graphicx}
\usepackage{apjfonts}
\usepackage{mathptmx}



\def\***#1{{\sc #1}}
\def\plan#1{\relax}
\def\Plan#1{\relax}
\def\PLAN#1{\relax}

\newcommand{\greq}{$\stackrel{>}{ _{\sim}}$}

\def\lta{\mathrel{\spose{\lower 3pt\hbox{$\mathchar"218$}}
     \raise 2.0pt\hbox{$\mathchar"13C$}}}
\def\gta{\mathrel{\spose{\lower 3pt\hbox{$\mathchar"218$}}
     \raise 2.0pt\hbox{$\mathchar"13E$}}}
\newcommand{\etal}{{\it et al. }}

\shorttitle{}
\shortauthors{}

\def\mathnew{\mathsurround=0pt}

\def\simov#1#2{\lower .5pt\vbox{\baselineskip0pt \lineskip-.5pt
\ialign{$\mathnew#1\hfil##\hfil$\crcr#2\crcr\sim\crcr}}}

\begin{document}

\title{The Unusual Spectrum of the Ultra Luminous X-ray Source, M82 X-1}

\author{ V. K. Agrawal\altaffilmark{1} and R. Misra\altaffilmark{2}
	 }

\affil{Inter-University Center for Astronomy and Astrophysics, Post Bag 4,
Ganeshkhind, Pune-411007, India}

\altaffiltext{1}{agrawal@iucaa,ernet.in}
\altaffiltext{2}{rmisra@iucaa,ernet.in}

\begin{abstract}
The results of a spectral analysis, using {\it XMM-Newton} and {\it Chandra}
data of the brightest
ultra luminous X-ray source in the nearby galaxy M82, are presented. 
The spectrum of M82 X-1, was found to be unusually hard 
( photon spectral index $\Gamma \approx 1$ ) with a sharp cutoff
at $\approx 6$ keV. Disk black body emission model requires
a  nonphysically high temperature. 
Instead, the spectrum is better described, with a lower reduced $\chi^2$, 
as emission
due to nearly saturated Comptonization of photons in an optically thick
($\tau \approx 10-30$, depending on the geometry) plasma having a 
temperature $kT \approx 2$ keV. This is in contrast to the high
energy spectra of other black hole systems, which are relatively
steeper ($\Gamma > 1.5$) and hence are modeled
as un-saturated thermal and/or non-thermal Comptonization of soft photons,
in an optically thin ($\tau \approx 1$) high temperature plasma. An Iron
line emission which is marginally resolved ($\sigma \sim 0.2$ keV) is
required to fit the data. 
We argue that the standard geometries for the X-ray producing region,
which are an optically thin inner disk or an uniform/patchy corona on top
of a cold disk, are not applicable to this source. Alternatively, the
geometry of the X-ray producing region could be a large sphere surrounding
a cold accretion disk or an optically thick inner disk region which
cools by bremsstrahlung self-Comptonization. For the latter
scenario, such an inner disk region, whose effective optical depth
to absorption is less than unity, is expected in the standard
accretion disk theory for near Eddington accretion rates.
\end{abstract}


\section{Introduction} 

{\it Chandra} observations of nearby galaxies
have detected several non-nuclear X-ray point sources, some of which
have isotropic luminosities $> 10^{39}$ ergs/s and are called 
Ultra Luminous X-ray sources, ULX \citep{Mat01,Zez02}. 
These results have confirmed and extended
the earlier evidence for the existence of such sources by the
{\it Einstein} survey \citep{Fab89, Fab87} and the 
ROSAT detection of  $\sim 100$ ULX in nearby galaxies \citep{Col02}.
Systematic studies of the {\it Chandra} observations
have revealed that ULX exist in  galaxies of different morphological
types. Since, by definition,
ULX emit radiation at a rate larger than the Eddington
luminosity for a ten-solar mass black hole, they are believed to
harbor a black hole of mass $50 \, M_\odot\! < \! M \! <\! 10^4
\,M_\odot$, where the upper limit is constrained by the fact that a
more massive black hole would have settled into the nucleus due to
dynamical friction \citep{Kaa01}. If this is true, then 
the differences and similarities
between the spectral properties of 
such intermediate mass black hole systems (IMBH) with those of
Galactic black holes (GBH) and Active Galactic Nuclei (AGN), 
are expected to enhance
our understanding of black hole systems in general. 
Alternatively, the ULX sources may be systems emitting at
super-Eddington luminosities \citep{Beg02}, or their emission may be
beamed from a geometrically thick accretion disk \citep{Rey97, Kin02}.
It has been argued that in the latter case, such thick
"funnel" shaped disks can enhance the observed flux by just a factor of
few \citep{Mis03}. In any case, the complexities involved in
explaining their origin \citep{Por02,Tan00,Mad01} and the inferred 
high accretion rates \citep{Kin01} 
make these enigmatic sources an important class of X-ray systems,
whose dominant radiative mechanism and geometry, needs to
be studied and constrained.

The spectra of ULX are expected to be different from GBH and AGN, if
the black hole is an IMBH and/or if their luminosity exceeds the
Eddington limit.
Earlier analysis of ASCA data revealed that the spectra of some ULX 
can be described as
radiation from a black-body emitting disk with an additional high energy 
power-law \citep{CM99},  
which is also the standard radiative model for  the high-soft 
 spectral state of Galactic black hole systems. 
The inferred inner disk temperature from these initial analysis, 
was found to be too high ($kT_{in} \sim 1.1-1.8 $ keV) for an accreting IMBH \citep{CM99,Mak00}.
However, recent analysis of 
{\it XMM-Newton} data have revealed the presence of a cool accretion
 disk component ($kT_{in} \sim 0.1-0.5 $ keV) in   NGC 1313 X-1, X-2
 \citep{Mil03} and M81 X-9 \citep{Mil04a} which
reaffirmed that these systems may indeed harbor an IMBH. The
spectra of some other ULX can be represented by a single power-law suggesting
a similarity with the low-hard state of GBH, where the disk emission
is either absent or weak. In fact, 
transitions between the two spectral states, similar to those seen in GBH, 
 have been reported in the ULX NGC 1313 X-1\citep{CM99} and two sources in IC342
\citep{Kub01}. Analysis of {\it Chandra} data
for a large number of ULX \citep[e.g.][]{Col04,Swa04} have revealed that
the spectra of most of these sources can be represented by a power-law with a 
typical photon index, $\Gamma \approx 1.6-2.0$. This spectral index 
is similar to
those of GBH in the hard state and Seyfert 1 AGN \citep{Zdz96a}. However,
most of these ULX are too faint for any further detailed spectral analysis.

One of the most intriguing ULX is the brightest X-ray source in  the 
starburst galaxy M82 that is located $\sim 170 pc$ away from the 
dynamical center. The isotropic luminosity of this source, M82 X-1, can
exceed 10$^{41}$ ergs s$^{-1}$ indicating that either it
is radiating well above the Eddington limit, or it harbors a black hole
$ > 100 M_\odot$, even if some moderate beaming (a factor of few) is taken
into account. ASCA observation have shown that this source is highly variable
with its luminosity in $0.5 - 10$ keV band changing from 
4.5$\times$ 10$^{40}$ to 1.6 $\times$ 10$^{41}$ ergs s$^{-1}$ \citep{PG99}.
The presence of a $54$~mHz QPO in its X-ray light curve suggests that
the source is a compact object and not highly beamed \citep{Str03}.
If this QPO is related to the  $0.8-3$ Hz QPO observed in galactic
sources, it would imply that its  
black hole mass, $M_{BH} \sim 100-300 M_\odot$.
Its high luminosity and relative proximity makes this source an
ideal ULX whose spectra can be analyzed in detail.

\cite{Swa04} have analyzed and fitted a large number of ULX observed
by {\it Chandra} including M82 X-1. After correcting for pileup,
they report that the 1999 {\it Chandra} observation of M82 X-1
  can be fitted by a power-law
with spectral index $\Gamma = 1.17^{+0.15}_{-0.14}$ and
$N_H = 0.98^{+0.12}_{-0.12} \times 10^{22}$ cm$^{-2}$. \cite{Str03}
mention that the {\it XMM-Newton} data can be fitted both by a disk black body
emission with an unphysically high temperature of $\sim 4$ keV
and a Comptonization model, but  they do not quantify the parameter 
values obtained nor discus the physical interpretation of the fits.
Recently, \cite{Fio04}, fitted the same {\it XMM-Newton} data with the bulk Comptonization
model and obtained a power-law index $\Gamma \sim 2$ indicating that
this source is similar to the soft/high state of black hole binaries.
However, as shown in this work, the column density required for such a fit, 
$N_H > 5 \times 10^{22}$ cm$^{-2}$, 
is not consistent with some of the {\it Chandra} observations.
Note, that the {\it Chandra} ACIS
data essentially measures the spectrum in the $0.5-5$  keV, since the
effective area of the detector decreases rapidly for energies greater
than $5$ keV. On the other hand, the {\it XMM-Newton} EPIC data measures the spectrum
in the $3-8$ keV range, since below $3$ keV the counts have contribution
from the diffuse background. This complementary nature of the two
types of observations makes it imperative to take into account both of
them in any spectral analysis. In particular the {\it Chandra} observations
give a better measure of the neutral column density, while the detailed
higher energy shape of the spectrum can be constrained by {\it XMM-Newton}
observations.  Since the source is known to be variable,
in principle,  any spectral analysis should be based only on simultaneous
observations, which currently do not exist. However, here we take a
pragmatic approach and use the non-simultaneous {\it Chandra} observations
of the source, to test whether the parameters obtained are consistent with
the results of the {\it XMM-Newton} analysis. 
Our motivation is to obtain the simplest physical radiative model 
for the source, which
in turn, will highlight any differences between this ULX and other
black hole systems.

\section{Observations and data reduction}

M82 was observed five times between September 1999 and June 2002 
by {\it Chandra}.  For three of these observations the
pileup fraction for M82 X-1, is $> 70$\% and hence 
no reliable analysis can be undertaken. 
For the other two ( whose observation ID numbers and
dates are tabulated in Table 1) the pileup fraction is around $25$\%, and
hence only these two were analyzed.
ACIS spectra of the source were extracted from circular regions 
of $\sim$ 2'' radius. 
Background spectra were extracted from nearby source free regions.
The analysis was done using the software, {\bf CIAO v3.2.1} and the
latest calibration files available till October 2005 were used to 
create the spectral response.

\begin{deluxetable} {lcccccc}
\tablewidth{0pc}
\tablecaption{Chandra Data Analysis}
\tablehead{
 \colhead{ObsID } & \colhead{Date}
& \colhead{$\Gamma$} & \colhead{$N_H$} & \colhead{C/s} & \colhead{L$_{39}$} & \colhead{PF (\%)}}
\startdata
1302 &1999-09-20 & 1.5$^{+0.1}_{-0.2}$ &1.2$^{+0.2}_{-0.2}$&0.12 &4.8 & 25  \\
2933 &2002-06-18 & 1.1$^{+0.1}_{-0.3}$ & 1.1$^{+0.2}_{-0.1}$ & 0.13 &5.0 & 26 \\
\enddata
\tablecomments{Chandra data observational ID, date of
observation, best fit power-law index ($\Gamma$), best fit column density
in $10^{22}$ cm$^{-2}$ ($N_H$), Count rate in the energy range 0.5-7.0 keV (C/s), 
estimated absorbed luminosity (using a distance  $D=3.9$ Mpc) for the same energy range 
in unit of $10^{39}$ ergs/s (L$_{39}$) and pileup fraction in percentage (PF)}
\end{deluxetable}

{\it XMM-Newton} observed M82 for a total elapsed time of $\sim 30$~ks on 
2001 May 5. Both the EPIC instruments PN and MOS were operated in 
Prime-Full-Window mode using the medium filter.
The PN events were processed using the {\it XMM-Newton}
analysis software ({\bf SAS v6.0}) and calibration data base 
available in August 2004. PN events corresponding to 
patterns 0-4 (single and double) were used for the spectral analysis.
The PN spectrum was extracted  using
a  circular region of $10{\arcsec}$ radius centered 
at the source position and the background spectrum was extracted from
a source free (based on the {\it Chandra} image) circular region.
The events were checked for the pile-up effect which was found not to be
significant. There are  two {\it Chandra} detected X-ray
point sources located less than $13{\arcsec}$ from M82 X-1. To check whether
these sources effect the analysis, we also extracted the PN spectrum
from a smaller $5{\arcsec}$ region and performed the same spectral fitting
as described in the next section, although the response for such a small
extraction radius may not be accurate. There were no qualitative difference in
the results of the spectral analysis.

\section{Spectral analysis}

Since the  Chandra observations are affected by pileup, the effect
was taken into account by using  the ``jdpileup'' model of the {\it
Chandra} spectral fitting software {\it Sherpa}. We fitted Chandra data
in the energy range 0.5-7.0 keV with ``jdpileup'' model
convolved with an absorbed powerlaw.  The best fit column density $N_H$,
the photon spectral index $\Gamma$, pileup fraction, count rate and
estimated luminosity are tabulated in Table 1. 
Since the pileup fraction is significant,
fitting with a more detailed model than an absorbed power-law is
not warranted. The qualitative inference that can be obtained from
this analysis is that column density $N_H \approx 10^{22}$ cm$^{-2}$
and that the spectra have a steep spectral index, $\Gamma < 2$. 

\begin{deluxetable}{lccccc}
\tablewidth{0pc}                                                                    
\tablecaption{Best fit model parameters for XMM data}
\tablehead{
\colhead{} 
              & \colhead{Diskbb}        & \colhead{CompTT}         & \colhead{CompTT}        & \colhead{Thcomp}       & \colhead{Bcomp}}
\startdata
nH            & 2.83$^{+0.51}_{-0.53}$    & 5.53$^{+0.46}_{-0.78}$   & 2.47$^{+2.12}_{-2.31}$ & 2.62$^{+1.42}_{-1.40}$ & 2.77$^{+1.40}_{-2.35}$ \\
              &                        &                         &                        &                        &                        \\
k$T_{in}$     & 3.32$^{+0.13}_{-0.10}$    &                         &                        &                        &                        \\
              &                        &                         &                        &                        &                        \\
kT            &                        & 35$^{+30}_{-11}$         & 2.24 $^{+0.28}_{-0.22}$& 2.34$^{+0.25}_{-0.12}$ & 2.46$^{+0.58}_{-0.34}$ \\
              &                        &                         &                        &                        &                        \\
$kT_s$        &                        & 0.30$^{+0.08}_{-0.27}$    & 0.30$^{+0.37}_{-0.28}$  & 0.13$^{+0.34}_{-0.12}$ &                        \\
              &                        &                         &                        &                        &                        \\
$\tau$        &                        & 1.16$^{+0.58}_{-0.30}$    & 11.10$^{+2.50}_{-2.40}$&                        &                        \\
              &                        &                         &                        &                        &                        \\
$\Gamma_{LE}$ &                        &                         &                        & 1.41$^{+0.15}_{-0.16}$ & 1.49$^{+0.23}_{-0.20}$ \\
              &                        &                         &                        &                        &                        \\
\hline
              &                        &                         &                        &                        &                        \\
$E_l$         & 6.64$^{+0.08}_{-0.13}$ &    6.4$^f$              & 6.67$^{+0.08}_{-0.13}$ & 6.67$^{+0.07}_{-0.07}$ & 6.67$^{+0.08}_{-0.13}$ \\
              &                        &                         &                        &                        &                        \\
$\sigma_l$    & 0.27$^{+0.11}_{-0.08}$ & 0.89$^{+0.32}_{-0.15}$     & 0.19$^{+0.11}_{-0.08}$ & 0.19$^{+0.15}_{-0.08}$ & 0.21$^{+0.15}_{-0.09}$ \\
              &                        &                         &                        &                        &                        \\
\hline
              &                        &                         &                        &                        &                        \\
$\frac{\chi^2}{dof}$  & 172/176                & 178/175                 &  171/174               & 171/174                & 171/175                \\
\enddata 
\tablecomments{Diskbb: Disk black body emission model. CompTT: Comptonization
model in Xspec. Thcomp: Comptonization model using the low energy spectral
index $\Gamma_{LE}$ instead of $\tau$ as a parameter. Bcomp: Bremsstrahlung
Self-Comptonization model. For all models a Gaussian line at $E_l$ and width
$\sigma_l$ has been included.\\
$^f$ $E_l$ fixed at 6.4 keV.  }
\end{deluxetable}

The {\it XMM-Newton} 
spectrum below $3$ keV is expected to be affected by the diffuse X-ray background, 
hence only  data in the energy range $3.0-10.0$ keV was used for spectral fitting.
Emission from a 
black body emitting disk, the XSPEC model ``diskbb'' \citep{Mit84}, 
does provide a
formal fit (Table 2), 
but 
the required maximum temperature of the disk
$T_{in} \sim 3.3$ keV is unphysical and would
require a large color factor correction $\sim 10$ for a disk radiating
at the Eddington limit around a $100 M_\odot$ black hole. Hence, a more
plausible explanation is that the spectrum is due to Comptonization.
We fit the data using the XSPEC model ``Comptt'' \citep{Tit94}, 
which describes Comptonization
of soft photons by a thermal distribution of electrons. There are two
different sets of spectral parameters which can describe the data.
The first set (second column of Table 2), has a high temperature, $kT \approx 35$ keV,
and low optical depth $\tau \approx 1$. Such a spectrum, which is due to unsaturated Comptonization
of external photons, is similar to the ones observed in other black hole systems and
 to the spectral fit obtained by \cite{Fio04}, using the  bulk Comptonization
model for the same data set. However, the required column density $nH \approx 5.5 \times 10^{22}$ cm$^{-2}$
is larger than the that estimated from {\it Chandra} observations (Table 1). The
second set of parameters (third column of Table 2), has a low temperature,
$kT \approx 2$ keV, and high optical depth $\tau \approx 10$. This spectra,
which is due to saturated Comptonization, gives a lower $\chi^2$  as compared
to the unsaturated case and the required 
column density (which has large error estimate) is similar to 
the {\it Chandra} value. An Iron line emission, which is required for both
the fits, turns out to be a narrow one at the expected centroid energy of
$\approx 6.7$ keV (for ionized Iron) for the saturated spectrum fit. On
the other hand, for the unsaturated spectrum, the required line energy
is unphysically small at $\approx 4.5$ keV and has a large width
of $\approx 2$ keV. Fixing the centroid energy to $6.4$ keV
(as reported in Table 1) still requires a large width of $\approx 1$ keV.
Thus, taking into account the smaller $\chi^2$ value, compatible column density
with {\it Chandra} observations and the simpler Iron line profile,
we consider the saturated Comptonization spectra as a better fit
to the data.

We confirm this result by also fitting the data
to the thermal Comptonization model described by \citep{Zdz96b}, named
``Thcomp'',  which takes
into account relativistic corrections (Table 2). Instead of the
optical depth $\tau$, ``Thcomp'' uses the geometry independent
low energy spectral index $\Gamma_{LE}$ as a parameter. The
optical depth can then be derived from  $\Gamma_{LE}$ and electron
temperature, giving $\tau \approx 12$ ($\approx 20$) for a slab (spherical)
geometry. The spectrum can also be interpreted as  emission from a thermal
plasma which Comptonizies internally produced bremsstrahlung 
photons, instead of external low energy ones. Such a  model, named ``Bcomp'', fits the
data equally well. Thus the fitting is
not sensitive to the spectral shape of the input seeds photons used
in the model.

\begin{figure}
\begin{center}
{\includegraphics[width=1.0\linewidth,angle=0]{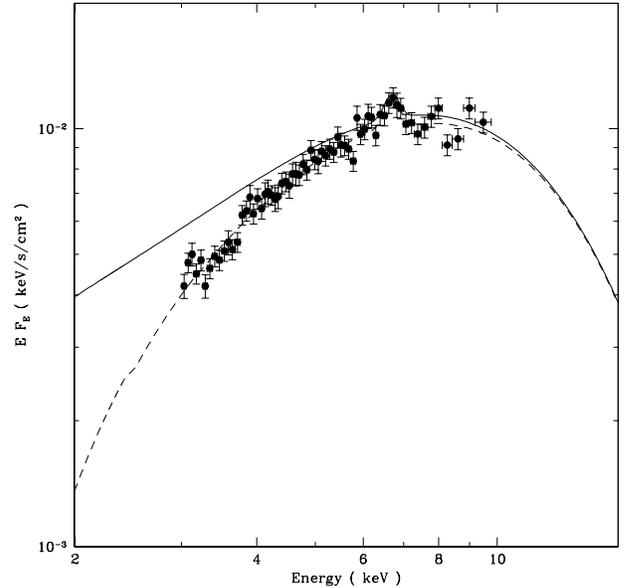}}
\end{center}  
\caption{ $E F_E$ unfolded spectrum of M82 X-1, for the soft
photon Comptonization model. The filled circles are {\it XMM-Newton}
data points which have been rebinned for clarity. The solid line is the
unabsorbed  model (while dashed line is the absorbed model) spectrum 
corresponding
to the best fit parameters given in Table 2.
 }
\end{figure}

The Compton Y-parameter ($y \equiv 4 kT \tau^2 /m_e c^2$) is a measure
of the degree of Comptonization that an average photon undergoes. From
the spectral fit parameters, $y \approx 3$, which implies that the
Comptonization process is nearly saturated. This is evident in
Figure 1, which shows the unfolded $E F_E$ spectrum for the  
Comptonization  model. The peak of the spectrum
is at $\approx 6$ keV $\approx 3 k T$, as expected for saturated
Comptonization. This is in contrast to un-saturated Comptonization
spectra which peak at $\approx kT$.
As reported by \cite{Str03} an Iron line feature is detectable
in the spectrum.

\section{Discussion}

The spectrum of M82 X-1 is unusually hard and hence 
the simplest radiative model which can describe it 
is saturated thermal Comptonization.
This is in contrast, to other black hole systems, where
the spectra is generally modeled as un-saturated thermal/non-thermal
Comptonization. The spectra of other ULX are steeper
with photon index, $\Gamma \sim 1.7$ \citep{Mil03,Mil04a,Mil04b}
similar to those of AGN and Galactic black hole systems. Thus
either, M82 X-1 is different from these systems or at least it
was in a rare and unique spectral state during the {\it XMM-Newton observation}.

The spectral fit parameters obtained
imply that the geometry of
the X-ray producing region for M82 X-1, is not the same as
that of other black hole systems.
The large optical depth $\tau$ \greq $10$  would prevent the 
entry of any external soft photons into the thermal plasma.
Hence, the standard geometry of
an hot inner disk, which Comptonizies photons from an outer cold
disk \citep{Sha76} is not applicable to this source. 
Similarly, soft photons will not be able to
enter an optically thick patchy corona or active region \citep{Pou99}
on top of a cold disk. On the other hand, the
soft photons from a cold accretion disk would diffuse out 
through an optically thick
corona if it uniformly covers the disk \citep{Lia77}. However, such models
are applicable only when the Compton amplification factor, $A \sim 1$.
Otherwise the Comptonized photons will impinge back on the disk
and heat the disk to the corona temperature \citep{Zdz96b}. Since
 for saturated Comptonization  the amplification factor
$A \approx T / T_s \approx 10$, the uniform corona model is
also ruled out. A possible geometry is that the inner region
of an accretion disk is surrounded uniformly by a thermal
plasma sphere. If the sphere is significantly larger than the
soft photon producing disk, only a small fraction of the
Comptonized photons will impinge back on the disk and energy
balance can then be maintained. Interestingly, the fitted
parameters suggest that the temperature of the accretion disk,
which can be identified as the soft photon source temperature,
$T_s \sim 0.3$ keV, is expected for a disk  around
a large $100 M_\odot$ black hole. However, it is not clear whether
such a large sphere would be dynamically stable or even realizable
in the presence of a black hole.

Another radiative model for the source
is that the thermal plasma Comptonizies photons that are 
internally produced by bremsstrahlung. 
In the standard accretion disk model \citep{Sha73},
the disk is assumed to be effectively optical thick to
absorption and hence the disk radiates as a black body locally.
However, at high accretion rates, the inner region of the disk may
become optically thin to absorption and the standard disk theory
is no longer consistent. This transition occurs roughly when
$\tau_* \equiv \sqrt{\tau_{es} \tau_{abs}} \approx 1$, which corresponds
to a transition radius
\begin{equation}
r_{tr} \approx 40 \; r_s \; \alpha^{34/93}\; \left (\frac{M}{100 M_\odot}\right )^{2/93} \left (\frac{\dot M}{\dot M_{Edd}}\right )^{64/93} 
\end{equation}
where $r_s \equiv 2GM/c^2$ is the Schwarzschild radius, $M$ is the
mass of the black hole, $\dot M$ is the accretion rate, 
$\dot M_{Edd} \equiv L_{Edd}/0.1 c^2$, is the accretion rate
corresponding to Eddington luminosity and  $\alpha$ is the 
dimensionless viscosity parameter.
For radii $< r_{tr}$, the disk cools inefficiently by 
bremsstrahlung Self Comptonization (BSC) and hence is hotter than 
a disk described by the standard theory. In this region, 
the temperature increases rapidly with decreasing radii. 
In certain cases, especially when the effectively optically thin region 
is an extended
one, the temperature can vary from $\approx 0.1$ keV at the
transition radius to nearly $\sim 50$ keV in the inner most radii. 
The resultant spectrum, which is a sum of local BSC spectra,
can have a power-law form and 
such a scenario has
been put forward as an alternate model to the standard unsaturated
Comptonization interpretation of the 
high energy spectra of AGN \citep{Mar90} and X-ray binaries \citep{Mis98}. 
Since the temperature increases inwards, the high energy photons are
preferentially produced in the inner region of the disk as compared
to the lower energy ones. Thus the observed time-lag between different
energy bands \citep{Now99} can be explained as waves traveling from the 
outer to the inner regions of the disk \citep{Mis00,Utt01}. 
For M82 X-1, the saturated
spectrum could arise from such an effectively optically thin region.
However, here the temperature variation is more modest, with a  maximum
of $\sim 2$ keV, indicating that the region is not an extended one.
Detailed spectral prediction from such a model depends on the
 actual temperature profile, and the present data is not adequate to
unambiguously constrain the physical characteristics of the region.
Nevertheless, it is interesting to note that M82 X-1 is probably
emitting close to its Eddington limit, even if it is a $100 M_\odot$
black hole, and at such high accretion rates, the inner region
of the standard disk is expected to become effectively optically thin.

The results obtained in this work need to be confirmed by
simultaneous {\it Chandra} and {\it XMM-Newton} observations. It will
also be interesting to find out using future {\it XMM-Newton} observations, whether
the source makes a transition to a state which has a regular steep X-ray
spectrum. Such an analysis will also shed light on whether or not the
temporal property of the source, especially the observed QPO, depends
on its unusual spectrum. Search and analysis of other ULX that have
similar hard spectra needs to be undertaken to verify whether M82 X-1
is indeed a unique source.

\clearpage

\end{document}